\documentclass[journal=ancac3,manuscript=article]{achemso}

 \usepackage[version=3]{mhchem} 
\usepackage{amsmath}
\usepackage{graphicx}
\usepackage[usenames]{color}
\usepackage[normalem]{ulem}
\usepackage{float}

\author{Felipe Bernal Arango}
\affiliation[AMOLF]
{Center for Nanophotonics, FOM Institute AMOLF, Science Park 104, 1098 XG Amsterdam, The~Netherlands}

\author{Andrej Kwadrin}
\affiliation[AMOLF]
{Center for Nanophotonics, FOM Institute AMOLF, Science Park 104, 1098 XG Amsterdam, The~Netherlands}

\author{A. Femius Koenderink}
\affiliation[AMOLF]
{Center for Nanophotonics, FOM Institute AMOLF, Science Park 104, 1098 XG Amsterdam, The~Netherlands}
\email{fkoenderink@amolf.nl}
\fax{+31 (0)20 754 7290}

 \title[Plasmonic Antennas Hybridized with Dielectric Waveguides]
  {Plasmonic Antennas Hybridized with Dielectric Waveguides}


\begin{document}

\begin{abstract}
For the purpose of using plasmonics in an integrated scheme where single emitters can be probed efficiently, we experimentally and theoretically study the scattering properties of single nano-rod gold antennas as well as antenna arrays placed on  one-dimensional dielectric silicon nitride waveguides. Using real space and Fourier microscopy correlated with waveguide transmission measurements,  we quantify the spectral properties, absolute strength and directivity of scattering. The scattering processes can be well understood in the framework of the physics of  dipolar objects placed on a planar layered environment with a waveguiding layer. We use the single plasmonic structures on top of the waveguide as dipolar building blocks for new types of antennas where the waveguide enhances the coupling between antenna elements. We report on waveguide hybridized Yagi-Uda antennas  which show directionality in out-coupling of guided modes as well as directionality for in-coupling  into the waveguide of localized excitations positioned at the feed element. These measurements together with simulations demonstrate that this system is ideal as a platform for plasmon quantum optics schemes as well as for fluorescence lab-on-chip applications.
\end{abstract}

\section{Introduction}


A highly promising development in nanophotonics is the use of plasmonic antennas to interface near fields and far fields\cite{Barnes2003,Novotny2011review,Aubry2010,Schuller2010}.  As opposed to conventional dielectric optics that are bound by the diffraction limit,  plasmonic structures can confine electromagnetic fields to very small volumes, essentially by packing energy in a joint resonance of the photon field and the free electrons in the metal.  As a consequence, plasmonic structures are currently viewed as ideal structures to interface single emitters and single photons \cite{Akimov2007, Curto2010,taminiau_08,Schietinger09,Falk2009,Carminati2006,bakker2006,Bolger2010,Rogobete:07,Agio2010,Koenderink2009,Pfeiffer2010}, as well as to realize nonlinear spectroscopies such as Raman enhancement\cite{Billot2006,Talley2005,Le2008,Ye:09}, SEIRA\cite{Jensen2000,Kundu2008,nakata2008,Le2008,Liu2010}, and so forth.\par
Currently, most workers in the field of nano-antennas target the basic understanding and use of antennas in essentially index-matched surroundings.  We propose that all the exciting properties of plasmonic nano-antennas can be used in even more versatile ways, if it would be possible to excite and interrogate the antennas efficiently in integrated photonic circuits.   Dielectric waveguides, such as high index ridges on low index substrates,  represent a common and mature photonic integration technology.\cite{Marcuse1974,Hill2004}  We envision local integration of plasmonic antennas as a promising route to enhance conventional dielectric photonic circuits, and  to achieve excitation and detection of plasmonic antenna resonances in an integrated fashion.
In order to ultimately apply this combination of structures it is important to understand exactly how antennas interact with waveguides, i.e., how the antenna scatters the waveguide modes, and conversely how the waveguide affects the antenna resonance frequencies, resonance profiles, and directivity. Antennas that are of particular recent interest are array antennas that consist of well understood individual objects, such as metal nanorods\cite{taminiau2007} with a strongly anisotropic polarizability, which  are placed in arrays of carefully engineered geometry\cite{Curto2010,hofmann07,Kosako2010,Aydin:10}. The physics of these systems is that electrodynamically retarded interactions set the strength and phase of coupling between elements, such that desired functionality ensues from interference.  For instance,  Yagi-Uda antennas\cite{Li2009,Koenderink2009,Waele2007,hofmann07,Kosako2010,Taminiau2008b} are phased array antennas that provide directionality to locally embedded fluorophores, due to constructive interference of the waves scattered by each antenna element in the forward direction.  While all the control variables in terms of building block size and shape, as well as the geometries that optimize interaction have been investigated by many researchers\cite{Kosako2010,hofmann07,Koenderink2009,Waele2007,Curto2010,Taminiau2008b,Offermans2011,Hao2009}, it is imperative to note that a strongly structured embedding dielectric environment will not only change the single building block response, but also the retarded interactions. Therefore, it is important to first study how single objects scatter when placed on waveguides, and subsequently to explore how array antennas function when placed on waveguides. A first step in this direction was taken in a recent publication\cite{Fevrier_12} by F\'evrier et al. who employed near field measurements to  show that the modes of arrays of gold scatterers on top of a silicon waveguide may hybridize\cite{Prodan2003} strongly with the waveguide mode.\par
In this paper, we present a comprehensive study of the different scattering channels that are available to a single plasmonic antenna that is placed on top of a dielectric waveguide. We show how we can understand small plasmonic antennas on top of a waveguide as dipolar scatterers in a layered system. We extend our theoretical model of a dipolar scatterer for the single-rod antenna to the multi-element antenna and find that the strong hybridization\cite{Prodan2003} of the antenna mode with the waveguide mode allows us to control the coupling of neighboring scatterers. We then use this understanding to design and experimentally study arrays of plasmonic scatterers with the purpose of maximizing the in-coupling of localized excitations, or the emission of single emitters into the waveguide. Lastly, we use a localized excitation based on a tightly focused spot, achieving high contrast efficient unidirectional coupling to the waveguide that can be controlled by wavelength.

\section{Results}
\begin{figure}[!htb]
  \includegraphics[width=8cm]{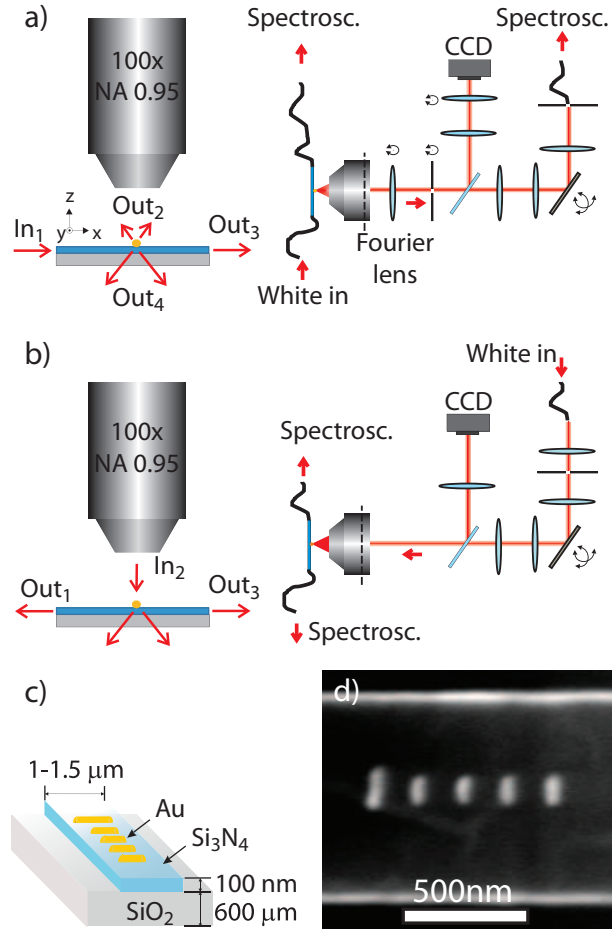}\\
  \caption{Schematic overview of the experimental setup [a) and b)] together with the representation of the two main working modes. In panel c) and d) the pictures present a schematic view of the sample used together with a scanning electron micrograph of a typical result of a fabricated Si$_3$N$_4$ waveguide with a deposited Au antenna.}\label{setupandsample}
\end{figure}
In order to study antennas coupled to dielectric waveguides, we employ a setup that combines a fiber-coupled end-fire setup with a confocal microscope as seen in figures \ref{setupandsample}a~and~\ref{setupandsample}b. The setup can be used in two configurations, as further highlighted in the sketches presented on the left side of the figures. In the first configuration of the setup shown in  Fig.\ref{setupandsample}a we send light from a supercontinuum laser into the waveguide using an input fiber, and we quantify the waveguide transmission spectrally, as well as performing microscopy on the out-of-plane scattering.  In the second configuration we do not input any light into the waveguide directly but instead study the reverse geometry. We quantify how light coming from free space is coupled into the waveguide depending on where, and with what wavelength, we excite the plasmon antenna. We  quantify the coupling by output that we pick up using fibers at the two waveguide end facets.  For a detailed description of all components in the set up, we refer to the Methods section. The samples used for the experiments are composed of gold antennas fabricated on top of silicon nitride waveguides by aligned electron beam lithography. A sketch of the structures used is shown in Fig.\ref{setupandsample}c. In this paper we discuss  two different types of antennas, namely single rod 100~nm long antennas and so called Yagi-Uda antennas~\cite{Kosako2010,hofmann07,Koenderink2009,Waele2007,Curto2010,Taminiau2008b}, where we have used the design of Curto et al.~\cite{Curto2010} but now coupled to waveguides. One such Yagi-Uda antenna defined over a Si$_{3}$N$_{4}$ waveguide is shown in Fig.\ref{setupandsample}d. While we have studied antennas on various waveguide widths, all the data presented here are for waveguide widths of 1000~nm and 1500~nm and strip heights of 100~nm. We estimate the electron beam alignment accuracy of antennas to waveguides to be $\sim$40~nm, i.e., far below any typical feature of the waveguide mode structure.\par

\subsection{Single element antennas on waveguides}
We first discuss measurements on single rod antennas excited through the 1000~nm width waveguides. These measurements are intended to obtain the resonance frequency of the single rod antennas. In order to obtain this information, TE polarized light is sent in through the waveguide (channel 1 in Fig.\ref{setupandsample}a), and light scattered by the antenna into the air-half space is collected and resolved on  the spectrometer (channel 2).
\begin{figure}[!htb]
  \includegraphics[width=8cm]{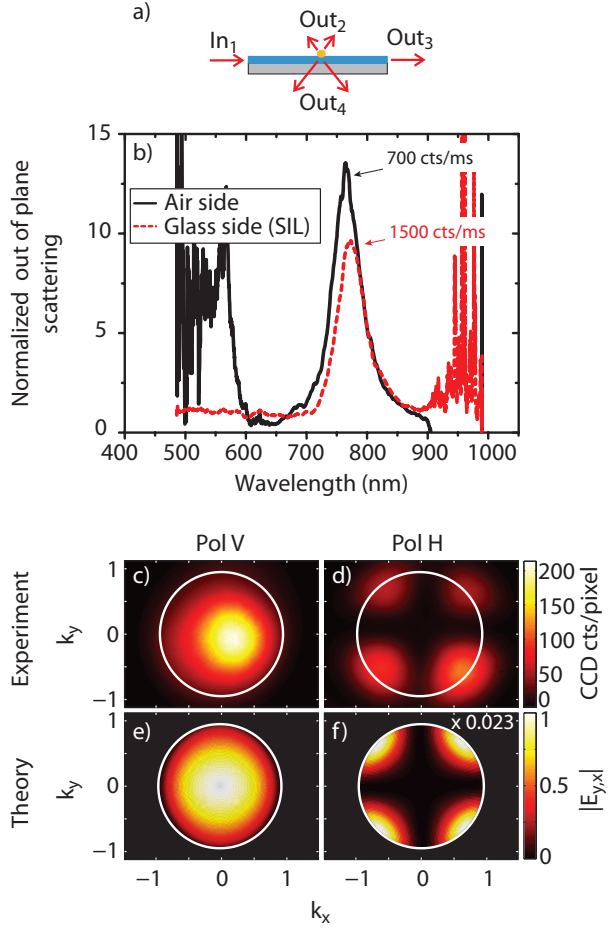}\\
  \caption{Sketch of the experimental geometry relevant for panels (b-f), in which we collect out-of plane scattering due to the antenna that is excited through the waveguide. b) Spectrum of the scattered intensity for a 100~nm long rod antenna on a 1000~nm width waveguide (continuous black line) taken from the air side of the sample. (dashed red line) taken from the glass side of the sample with a SIL. The spectrum is normalized to the light offered to the antenna, measured by integrating light scattered from roughness of the waveguide adjacent to the antenna. Both peaks show that the scattering of guided modes happens through a resonant process. c)and d) Graphs of the measured radiation pattern for the 100~nm rod antenna, analyzed through a vertical c) and horizontal d) linear polarizer. The white circle indicates the NA of the Olympus objective (N.A=0.95). The integration time for the vertical polarization is 1.45~s and 30~s for the horizontal polarization.  e) and f) Graphs of the simulated radiation pattern analyzed through a vertical e) and horizontal f) linear polarizer. Field values are calculated at the position of the microscope objective, i.e., 1.8~mm from the sample plane. The fields are normalized to $\mathbf{E}_{y}=1.7\times10^{-9}$~V/m, given a guided mode strength of 1~V/m. Graphs c) to f) demonstrate that a 100~nm rod antenna located over a multi-layer substrate behaves as an electric dipolar scatterer.}\label{singleobjectALLpart1}
\end{figure}
Fig.~\ref{singleobjectALLpart1}b (continuous black line) shows the   spectrum of light scattered into the air side of the sample by a  100~nm rod antenna normalized to the input  intensity with which it is excited through the waveguide according to the normalization method presented in the  Methods section.  The antenna spectrum shows a clear peak centered around 750~nm, with a bandwidth of around 55~nm (FWHM). The resonance frequency is comparable to resonance frequencies previously found for rods on simple glass substrates\cite{Moussa2008,slaughter2010,kalkbrenner04,Lindfors2004}.  When collecting light scattered by the antenna into the substrate underlying the waveguides using the (SIL) solid immersion lens system (see Methods section on microscope setup), we find that the resonance frequency is almost identical (Fig.\ref{singleobjectALLpart1}b (dashed red line)). However, when comparing the intensity of the light emitted into the different media for quantitative reference, we find that a signal approximately 2 times stronger is found into the quartz substrate than into air, consistent with the fact that a higher scattering intensity towards the high index medium is expected from the radiation of  dipoles on top of a high index layered system\cite{Novotny06b}.
We conclude from our  measurements  that waveguide-addressing of plasmon antennas allows for  high signal-to-noise ratio   dark-field spectroscopy of single plasmon antennas both using collection of light from the air side, and from the substrate side. This conclusion is highly promising for integrated applications of plasmonic antennas in sensing using integrated optics.  Even more promising is that detection in such a sensing scheme could also occur via the waveguide itself. Indeed, we estimate that a single nanorod antenna removes approximately 20\% of the intensity in the waveguide mode out of the transmission channel [data not shown] and redistributes it over waveguide reflection, absorption in the metal, and out-of-plane scattering.  This estimate results from transmission spectra normalized to nominally identical blank waveguides. In the remainder of this paper we focus quantitatively on this redistribution for single antenna elements, and explore how it can be controlled using  multi-element antennas.

To obtain a more comprehensive understanding of how plasmon antennas scatter waveguide modes, we  analyze the scattered light further in terms of polarization and directionality.  In the remainder of this paper we focus on collection of light on the air side of the sample, as the quality of our imaging system is far superior in this configuration.  Polarization analysis shows that more than 90\% of the light is scattered in the polarization direction parallel to the direction of the antenna (y direction in the reference frame depicted in Fig.\ref{setupandsample}a).   We have also studied 100~nm rod antennas fabricated at various rotation angles relative to the waveguide axis. As the antenna is rotated from  $90^{\circ}$,$45^{\circ}$ to $0^{\circ}$ angle relative to the waveguide axis,  we consistently find strong polarization of scattered light collected on the air side of the sample along the  antennas axis. We have access to the directionality of  scattering by  the single rod antennas using Fourier microscopy, i.e., by insertion of a  Bertrand lens into our imaging system.  While the 100~nm line rod antennas appear as diffraction limited points in spatial imaging, interesting information is obtained when looking at the scattered light by imaging the back focal plane of the objective in this manner. At the air side  (channel 2 in Fig.\ref{setupandsample}a ), the radiated pattern  appears to be distributed over a wide range of angles (up to $\sin\theta=0.7$) relative to the sample normal (Fig.\ref{singleobjectALLpart1}c~and~\ref{singleobjectALLpart1}d). Upon polarization analysis with a linear polarizer in detection we find a large contrast in integrated intensity. In addition, the weak cross-polarized radiation pattern is clearly distinct from the co-polarized pattern in that it consists of four separate lobes. Similar results were reported in\cite{sersic2011} for antennas excited using total internal reflection on a prism. Clearly, waveguide excitation is an efficient alternative to TIR for dark-field Fourier microscopy. We note that the fact that the radiation pattern extends somewhat \emph{outside} the NA of our objective indicates that diffraction by the spatial selection pinhole blurs the measured radiation pattern.  The radiation pattern found in the two polarization channels for the 100~nm rod antenna in Fig.\ref{singleobjectALLpart1}c~and~\ref{singleobjectALLpart1}d bears the clear signature of an in-plane  y-oriented dipole placed on top of a Si$_{3}$N$_{4}$-SiO$_{2}$ substrate.  In high-NA imaging, such a y-oriented dipole generates cross polarized fields at very large angles due to the huge refraction angles in the aplanatic imaging system. The measured radiation pattern  for the 100~nm rod antenna, as shown in Fig.\ref{singleobjectALLpart1}c~and~\ref{singleobjectALLpart1}d is in excellent agreement with the theoretical radiation pattern that is expected for  a dipolar scatterer positioned 30~nm above a 2D layer system consisting of quartz and siliconnitride as shown in Fig.\ref{singleobjectALLpart1}e~and~\ref{singleobjectALLpart1}f.
These theoretical  radiation patterns are calculated by using the analytically known far-field expansion of the Green's function of a multilayered system, as explained in ref.\cite{Novotny06b}. In this type of calculation the substrate is an infinitely extended multilayered system composed of Air-Si$_{3}$N$_{4}$-SiO$_{2}$, thereby ignoring the finite width of the 1D waveguide on SiO$_2$. In this 2D waveguide geometry, we can perform quantitative scattering calculations for arbitrary collections of antenna particles excited by waveguide modes that we obtain  by solving for the propagation constants and mode fields of guided modes that are bound to the Si$_{3}$N$_{4}$ waveguide.
In this approach, the polarizability of  scatterers is taken as the electrodynamically corrected quasi-static polarizability of a prolate spheroid\cite{bohren83}. The dynamical correction used in the calculations for the polarizability is\cite{Coenen2011} 
\begin{equation}\label{correctedpolarizability}
\overleftrightarrow{\alpha}^{-1}=\frac{1}{\overleftrightarrow{\alpha}_{static}}\overleftrightarrow{I}-(Im(\overleftrightarrow{G}_{scatt}(r_{0},r_{0}))+Im(\overleftrightarrow{G}_{0}(r_{0},r_{0}))).
\end{equation}
Here $\alpha_{static}$ is the static polarizability of a prolate spheroid, $\overleftrightarrow{G}_{scatt}(r_{0},r_{0})$ is the Green's function of the layered system as found in\cite{Novotny97a} evaluated at the position of the scatterer $r_{0}$; finally $Im(\overleftrightarrow{G}_{0}(r_{0},r_{0})))$ is the imaginary part of the free space Green's function. The satisfactory correspondence between the measured radiation patterns  for antennas on 1D guides, and the theoretical figures for 2D guides implies that  the finite width of 1~$\mu$m of the waveguide used does not strongly alter  the angular distribution of light scattered out-of-plane. Arguably, close inspection of the data shows that  angular emission  is narrowed in $k_y$ by the 1D waveguide compared to the 2D system.\par
\subsection{In-coupling by a single dipole antenna}
\begin{figure}[!htb]
  \includegraphics[width=8cm]{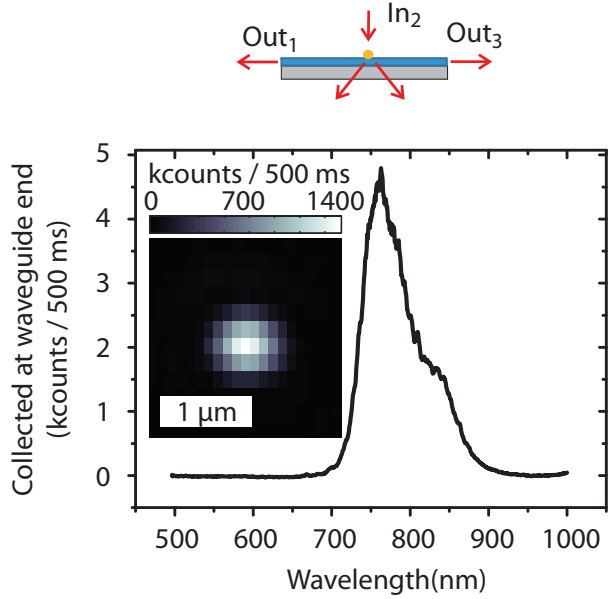}\\
  \caption{ Top: sketch of the experimental geometry. We excite the antenna from the air side, and measure how much light is coupled into the waveguide. Bottom: spectrum of the light collected at the waveguide end facets, i.e., of the light coupled into the waveguide upon excitation of the scatterer. A clear plasmon resonance is observed. In the inset we show a confocal raster scanning graph of in-coupled intensity for different positions of the focused spot.}\label{singleobjectALLpart2}
\end{figure}

As a complementary experiment  on the antenna-waveguide system, we have also performed the reverse, i.e., excitation from the far field and detection through the waveguide (see Fig.\ref{setupandsample}b). In this experiment a diffraction limited focused spot is scanned over the antenna and the light in-coupled into the 1500~nm wide waveguide is acquired through the aligned optical fibers at the waveguide end facets. The inset of Fig.~\ref{singleobjectALLpart2} shows a plot of the maximum in-coupled intensity for different positions of the scanned beam. The 2D color plot, which could be viewed as a confocal raster scanning graph, barring the fact that collection is through the waveguide, and not through any objective, indicates that the light is being coupled into the waveguide from a point that is approximately equal in size, or less, than the diffraction limit. By calibration of the spot to a white light image of the fabrication markers, we ensured that the center of the maximum in-coupled intensity coincides  strictly with the antenna position. At each position we furthermore collect spectral information, as the incident beam has a broad spectrum and the detected light is coupled into the spectrometer. Fig.~\ref{singleobjectALLpart2}  shows the spectrum at the location of maximum in-coupling determined from the 2D spatial raster scan. We find a maximum coupling from free space into the waveguide at a wavelength around 750~nm across a bandwidth of 65~nm (FWHM). The excellent correspondence of the in-coupling resonance frequency with the scattering resonance we observe when illuminating through the waveguide, indicates that resonant in-coupling into the waveguide occurs at the same wavelength as scattering of the waveguide mode by the antenna out of the waveguide. Also, the bandwidth agrees with the measured bandwidth in the out-coupling experiment. However, the spectrum in the in-coupling experiment has a tail towards the near infrared wavelengths due to the red shifted cutoff frequency of the 1500~nm wide waveguides compared to the 1000~nm wide waveguides used to obtain Fig.~\ref{singleobjectALLpart1}.\par
\begin{figure}[htb]
  \includegraphics[width=8cm]{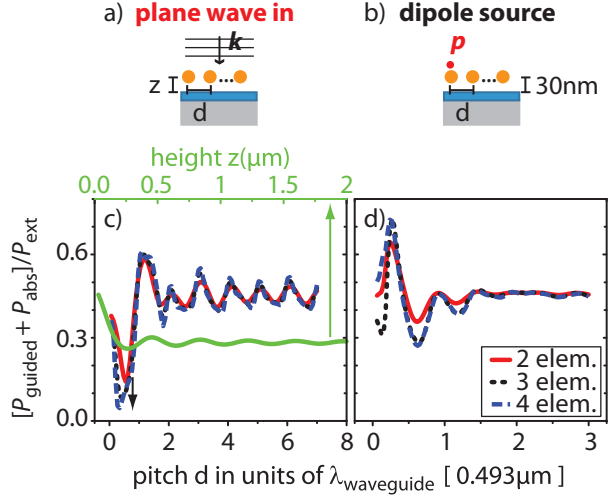}\\
  \caption{a) Presents a sketch of the calculations shown in panel c). In these calculation a plane wave is sent towards the antenna from the air side and the efficiency of absorption plus scattering into the waveguide mode is calculated. c) (green line) ratio between guided plus absorbed power to extinction power for plane wave excitation of the antenna found for a single rod antenna element at different `z' distances from the waveguide at 755~nm. (blue, black and red line) Using a height of 30~nm multi-element antennas (2,3 and 4 elements) are investigated for different distances `d' between elements composed of 100~nm Au rod antennas. b) Presents a sketch of the calculations shown in panel d) where an emitter is positioned 30~nm above the first element of an antenna to calculate the absorption efficiency plus emission efficiency of the antenna into the guided mode. This ratio of in-coupled and absorbed light to extinct power is presented for multi-element antennas (2, 3 and 4 elements) as a function of the distance between the elements when the antenna is being excited with the dipolar emitter located above the first element of the array. Horizontal axes are in units of $2\pi/\beta=0.493\mu$m which is the wavelength of the guided mode at 755~nm.}\label{theoryincouplingAll}
\end{figure}

We now attempt to estimate the in-coupling efficiency of light into the waveguide from the data measured in Fig.~\ref{singleobjectALLpart2}. In this experiment  coupling from the waveguide to the spectrometer used a metallized tapered fiber tip at the waveguide end facets, in order to reduce stray light contributions such as grazing light coupled to the SiO$_2$ substrate. Unfortunately, the use of this metallic tip makes it difficult to find a quantitative  coupling efficiency of antenna to waveguide, as the waveguide-to-fiber efficiency is imprecisely known.  On basis of in-coupling intensity data of 10 kcts/s at 750~nm, knowing that the irradiance factor for our spectrometer is 12.86~(kcts/s)/($\mu$W/cm$^{2}$/ nm) at 750~nm, we can calculate an in-coupled irradiance of 0.77 $\mu$W/(cm$^{2}\cdot$nm) in the spectrometer; when using a free space focused beam with an irradiance of 1.56$\times$10$^{8}\mu$W/(cm$^{2}\cdot$nm), an estimated fiber collection efficiency of $10^{-4}$, an efficiency in the single mode to multi mode fiber coupling of 10\%, and a loss in the waveguide of $10^{-2}$. With this data we estimate in-coupling efficiencies on the order of 1\%,  for diffraction limited in-coupling beams.
Unfortunately, the experimental uncertainties especially regarding the in-coupling of the signal into the detection fibers, imply that our experimental estimate is not more accurate than approximately one order of magnitude. To obtain an independent, and possibly more precise estimate, we turn to theory. We use the model of a dipolar scatterer on top of a 2D waveguide as explained before. We find the efficiency with which such a scatterer couples light into the waveguide in two steps. First, we find the extinct power, i.e., the power that is removed from a plane wave incident from the air due to the presence of the scatterer. The extinct power is defined as
\begin{equation}\label{extinction}
P_{ext}=\frac{\omega}{2}Im( \vec{p}\cdot\vec{E_{o}^{*}}).
\end{equation}
Furthermore, the power that the induced dipole moment radiates into the far field \emph{barring} the waveguide mode can be calculated from the dyadic Green's function far field expansion that can be found in Ref.\cite{Novotny06b}.  The difference in extinct power and far field radiated power equals the power coupled into the waveguide, plus the power absorbed by the particle due to losses.   We find (assuming a plane wave excitation) that the coupling efficiency strongly depends on the height of the single rod antenna with respect to the waveguide as  shown in Fig.\ref{theoryincouplingAll}c (green curve). This dependence reflects the strong spatial dependence of both the guided mode contribution, and radiative mode contribution to the local density of states of stratified waveguide systems. We predict a maximum incoupling+absorption of $\sim$48\% for particle heights 30~nm from the waveguide. This in-coupling decays exponentially with  distance from the waveguide and stabilizes at 30\% at distances around 2~$\mu$m from the waveguide. Since significant in-coupling is not expected for such large distances we estimate that those 30\% correspond to absorption in the particle.   Taking that as a measure for absorption, we conclude that a particle just above the waveguide will couple approximately 20\% of the light that it harvests from the input beam into the waveguide.  The remaining 80\% is split between far field (50\% of extinct power)  and absorption (30\% of extinct power). It is important to notice that these numbers indicate the efficiencies with which the power is distributed in the different radiation channels relative to the total power that couples to the dipolar scatterer. To convert these relative efficiencies to actual cross sections, one needs to determine what the absolute extinction cross section of the particle is. The overall extinction cross section is anticipated to be at most 0.16~$\mu$m$^2$, i.e., 1.2 times smaller than the diffraction limit. To conclude, a single particle illuminated by a diffraction limited beam can couple approximately 20\% of the incident energy into the waveguide, in accord with the crude measured estimate.\par
The constraint of fairly large absorption (30\%), which in our system is due both to the gold and to the underlying Cr adhesive layer, can be mitigated by shifting the operation range further to the NIR using larger particles, or by swapping Au for silver. In this case a protective dielectric could be required to avoid particle degradation. Such capping is expected to also be beneficial optically, as it would pull the waveguide mode up towards the particle, thereby likely enhancing the coupling efficiency.
\subsection{Multi-element antennas}
We now turn to multi-element antennas, where we can apply our understanding of the operation of single-element antennas to improve the absolute in-coupling cross section, albedo, and directivity, similar to the functionality of free space Yagi-Uda antennas\cite{Curto2010,Kosako2010,Koenderink2009,Taminiau2008b}. An important realization is that if we have a dielectric system that consists of planar waveguides or 1D waveguides, understanding coupled systems is a two step process.  Firstly,  the induced dipole moments in $N$ particles will be set by\cite{Novotny06b}
$$\mathbf{p}_{n} =\alpha [ \mathbf{E}_{\mathrm{in}} (\mathbf{r}_n) + \sum_{m\neq n} \mathbf{G}(\mathbf{r}_m,\mathbf{r}_n)\mathbf{p}_{m}],$$
where  the driving field $\mathbf{E}_{\mathrm{in}} (\mathbf{r}_n)$ is a solution to the antenna-free problem, such as a waveguide mode, or far field illumination. The Green's function  $\mathbf{G}(\mathbf{r}_m,\mathbf{r}_n)$ of the waveguide system quantifies the particle interactions as they are mediated through waveguide, substrate, and air cladding layer.  The second step in understanding the physics of multi-element antennas is that the near fields, far fields, etcetera,  are found by coherent superposition of the single-element properties as\cite{Novotny06b}
$$\mathbf{E}(\mathbf{r})= \sum_{n} \mathbf{G}(\mathbf{r},\mathbf{r}_n)\mathbf{p}_{n}.$$
It is this second step, for which we have quantified properties above, that ensures that multi-element antennas can control directivity, albedo, etc., just as for antennas in free space\cite{Curto2010,Kosako2010,Koenderink2009,Taminiau2008b}. The linear superposition principle will, for instance, imply that the radiation pattern into free space and waveguide of a multi-element antenna equals that of a single-element antenna (form factor) \emph{multiplied} with a structure factor that depends on where the different elements are placed. As a consequence, light can never be redirected into directions into which the single elements do not radiate, but light can be significantly redistributed through interference between the different channels into which the single elements do radiate. Thus, one can for instance seek to obtain enhanced radiation into the waveguide and suppression of radiation into substrate and air, through destructive interference.\par
Given that plasmon particles couple strongly to each other both directly and through  coupling mediated by the waveguide one can design multi-element antennas with different final purposes. One design goal is to achieve antennas that maximize the coupling of incident plane waves into the waveguide. Another design goal is to achieve an antenna which maximally couples energy from a single dipolar emitter into the guided mode. The latter would essentially constitute a waveguide-coupled plasmon Yagi-Uda antenna. Here we consider both design goals. First we focus on optimum structures for coupling plane wave excitation incident from the air side into the waveguide using the dipolar antenna building block at fixed height. As optimization parameter, we scan the distance between elements and evaluate the coupling, as shown in Fig.\ref{theoryincouplingAll}c.  On the x-axis we plot the distance between elements in units of the guided mode wavelength $\lambda_{waveguide}$ at 755~nm. We find maximum in-coupling at distances which are $n$ times $\lambda_{waveguide}$ (with $n$ an integer) and minimum when the distance is $(n+1/2)$ times $\lambda_{waveguide}$. Since the scatterers are driven in-phase, the arrangement essentially reflects that just a few particles already result in the well-known effect of a grating coupler, that can boost incoupling by a factor $\sim$2 to 3.\par
As a second example, more appropriate for extending plasmon quantum optics to waveguide integrated systems, we consider the scenario of a waveguide coupled Yagi-Uda antenna.  Here, the design goal is to couple the radiation of a single dipolar emitter, such as a localized molecule, quantum dot or diamond NV center selectively and unidirectionally to a single waveguide mode.  The design goal is hence for the radiation of antenna elements and dipole to add up destructively everywhere, except in the waveguide. In this case, a dipolar emitter is located 30~nm above the first element of the antenna. This emitter generates the driving field $\mathbf{E}_{\mathrm{in}} (\mathbf{r}_n)$ over the $n$ elements of the antenna. With this field we calculate the induce dipolar moments of the antenna elements which subsequently are used to find the scattered field, as explained earlier. In Fig.~\ref{theoryincouplingAll}d we plot the incoupling+absorption rate, as a function of the distance between the directors in the antenna array. As in figure \ref{theoryincouplingAll}c we plot distance in units of the waveguide mode propagation wavelength. A maximum in-coupling is found for a range of separation distances centered around $\sim\lambda_{waveguide}/4$ at 755~nm and ranging from $\sim$0.1 to $\sim$0.45~$\lambda_{waveguide}$. This range is commensurate with the standard rule of thumb for free space Yagi-Uda antennas, that the spacing needs to be around $\lambda$/3,  and  below $\lambda$/2 to avoid multiple lobes. However, in this case the criterion uses the wavelength of the waveguide mode. The optimum design hence depends on the dispersion of the waveguide. As regards in-coupling efficiencies, this calculation predicts that less than 20\% of the emission is directed out-of-plane into either air or substrate. At the same time, 80\% of the emission that is captured by the waveguide will be directed in a narrow forward lobe, with an angular distribution in a half angle cone of 37$^{\circ}$ inside the 2D Si$_{3}$N$_{4}$ layer. We note that Yagi-Uda antennas realized sofar have been essentially free-space designs placed for practical reasons on  air/glass interfaces \cite{Curto2010,Kosako2010,hofmann07}. In this scenario, a directional beam results, that is however completely directed into the glass, along the critical angle for the air-glass interface\cite{Taminiau2008b}. The utility of this form of directionality for on-chip applications is very limited, as the directional beam is directed out-of-plane. Here we show that this limitation can be entirely overcome by placing the  Yagi-Uda concept into, or on a waveguiding dielectric structure. Our calculation shows that even moderate waveguide confinement strongly influences the directionality to be entirely in-plane and into the guided mode. Thereby, the Yagi-Uda-waveguide combination could be a promising route to plasmon quantum optics. As opposed to, e.g., the plasmonic nanowire paradigm\cite{chang2006,Akimov2007,Falk2009} that foresees quantum optics in networks in which the excitation remains in dark plasmons throughout, here the utility would be that flying qubits experience entirely lossless transport through established dielectric technology, and conversion to plasmons for light-matter interaction is localized to the sites where it is needed.\par
\subsection{Measurements of waveguide excited multi-element antennas}
\begin{figure}[H]
  \includegraphics[width=16cm]{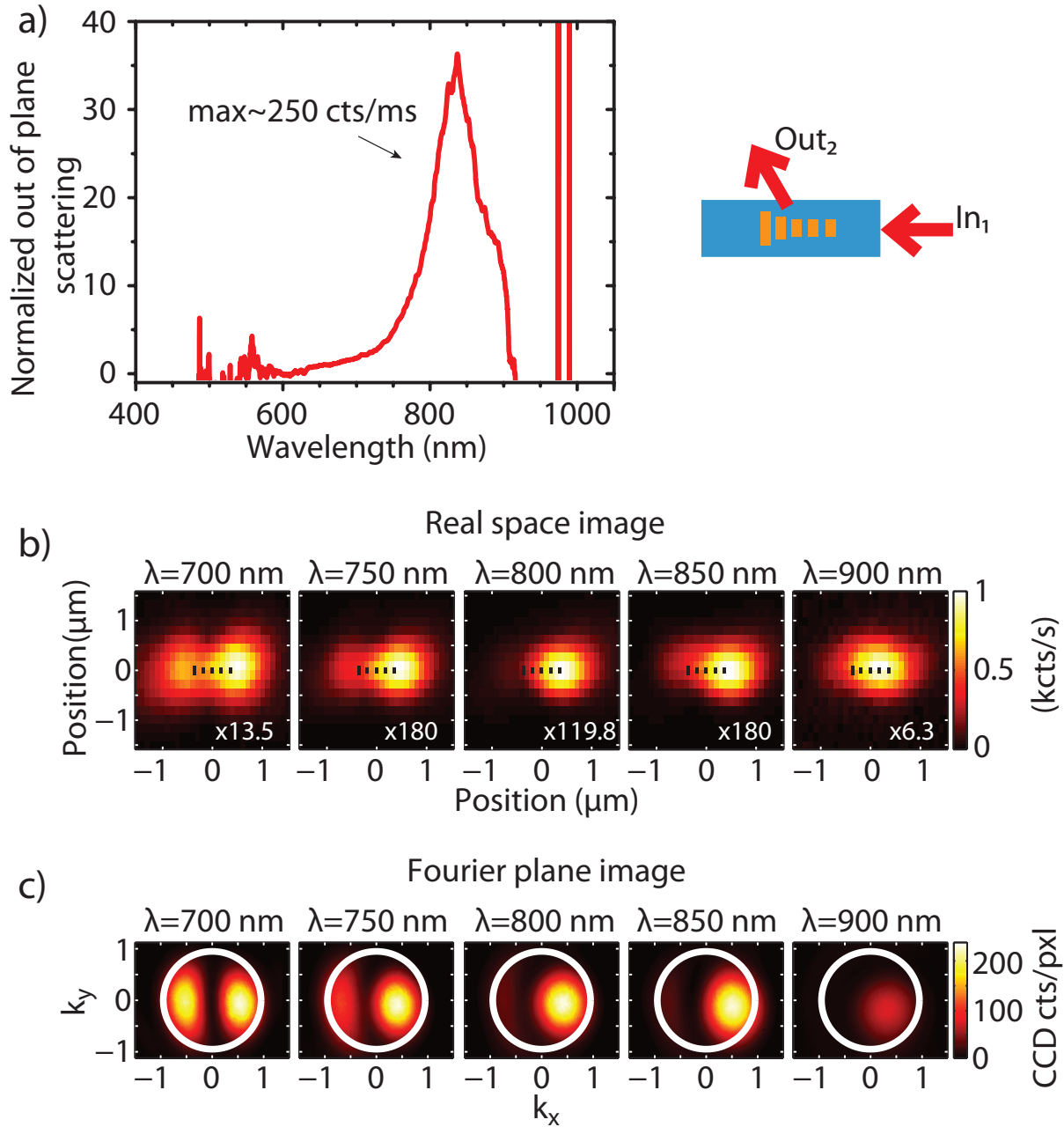}\\
  \caption{
  This figure shows measurements of the scattering of the guided mode to free space modes carried out on a Yagi-Uda antenna on top of a 1000~nm width waveguide.
   a) Normalized spectrum of the scattering into the free space of the guided mode by the antenna.
    b) Confocal scans over the sample plane of the scattering due to the antenna. As sketched in the cartoons on the top right side of the graph, the reflector is located at the left side of the image and the directors are located at the right side of the image. The beam is incident from the right side through channel~1 and the scattering out-of-plane is measured in the air side through channel~2. The graphs in b) show different frequencies measured when focusing at different positions in the sample plane. c) Fourier images of the Yagi-Uda antenna for the scattered light coming from the guided modes. The images are acquired with a CCD camera and the different ranges of wavelength are selected using bandpass filters with 40~nm FWHM. These graphs show that the antenna on the substrate has directionality in its scattering. (The integration times are: for 700~nm 9.64~s, for 750~nm 3.09~s for 800~nm 1.45~s, for 850~nm 6.6~s and for 900~nm 30~s.)}\label{YUmultielementAllPartA}
\end{figure}

The Yagi-Uda antennas that we fabricated on top of the Si$_{3}$N$_{4}$ waveguides have a director periodicity of 183~nm or $0.32\cdot\lambda_{waveguide}$ at antenna resonance (vacuum wavelength 830~nm),  well in the range predicted to provide directional behavior according to the theory of the previous section (Fig.\ref{theoryincouplingAll}d). We investigated the Yagi-Uda antennas using the same methodology applied to the single-object nano-antennas. The measurements are shown in Fig.\ref{YUmultielementAllPartA}. The spectrum of the guided mode scattering  into the air side of the sample shows a resonance at a wavelength of 830~nm, as shown in Fig.\ref{YUmultielementAllPartA}a. This resonance is significantly redshifted compared to the resonance of single plasmon particles in Figure~\ref{singleobjectALLpart1}. In part this shift is due to the fact that the antenna elements are slightly longer than the 100 nm rods, causing a redshift of the resonance in each particle. In another part, this shift is due to the fact that the antenna response of Yagi Uda antennas is red shifted by plasmon hybridization\cite{Prodan2003}, as reported already by~\cite{Koenderink2009,Waele2007}.
Quite unlike is the case for a single rod antenna, the spectrum strongly depends on the part of the antenna from which the light is detected by our imaging system (see Fig.\ref{YUmultielementAllPartA}b). In fact in a spectrally resolved raster scan of the antenna one can visualize two clearly distinct zones which change relative intensity depending on the wavelength. Naively one might assume that a confocal scan reports an image of the local field intensity $|E|^{2}$ at the antenna, blurred by the diffraction limit. In this view, the appearance of distinct bright zones at different wavelengths would indicate spatial localization of induced dipole moments $|p|^{2}$ along the chain, similar to the report by de Waele et al.\cite{Waele2007}. In reality, more information is hidden in our data, since image formation is a coherent process that actually results from interference of radiation from all the dipoles in the sample plane on the confocal pinhole. Thus, phase information is also hidden in the confocal images, and the collected spatial distribution should not be interpreted simply as a map of $|p|^{2}$ with Gaussian blurring due to the diffraction limit. We have performed calculations [Not shown] that include the interference in the image formation process using the amplitude and vector microscope spread function of confocal microscopy\cite{Novotny06b}, similar to the calculations used to support the measurement of wavelength-tunable localization of dipole excitations on plasmon chains in index-matched environments reported in Ref.\cite{Waele2007}. Our calculations confirm that simply changing the input wavelength strongly changes both the spatial distribution of induced dipole strengths, as well as the distribution of phases excited along the array. For instance, having all elements in phase results in the antenna appearing as a bright entity in the confocal scan (Fig.~\ref{YUmultielementAllPartA}b, 900~nm wavelength). Conversely, for wavelengths where the antenna lights up as two distinct regions (Fig.~\ref{YUmultielementAllPartA}b, $\lambda<800$~nm), a 180$^{\circ}$ phase jump occurs in the induced dipole moments along the length of the plasmon chain. To conclude, the spatial maps prove that the Yagi-Uda antenna indeed acts as a phased array, driven coherently by the waveguide mode.\par

The spatial mapping shows that, depending on excitation wavelength, the amplitude and phase of the dipole excitations on the particle chain is strongly varying.  It is exactly this physics that gives rise to the interference that makes a Yagi-Uda antenna directional. Indeed, an excellent way to assess the coupling between antenna particles is to map radiation patterns for different driving conditions. We measure the radiation patterns on the air side using our Fourier plane imaging. We select distinct wavelength slices using 40~nm bandwidth band pass filters placed in front of the CCD camera. The scattering into the air side shows strong directionality with a distinct wavelength dependence (Fig.~\ref{YUmultielementAllPartA}c). From these measurements we see that Yagi-Uda antennas, when placed over a waveguide system, continue presenting directionality in their scattering. At 850~nm close to the scattering resonance the scattering in the forward direction is maximum. Far from resonance at 700~nm the backward directed scattering and the forward directed scattering have the same intensity. The scattered light is highly polarized in the direction parallel to the antenna elements, with polarization ratio > 1:9. In conclusion,  Yagi-Uda antennas on top of waveguides allow spectrally controllable directional out-coupling of waveguide modes, as well as wavelength and excitation direction dependent control of amplitude and phase along the length of the antenna,  very similar to observations recently made in scattering experiments\cite{Koenderink2009,Waele2007}. Such tunable radiation patterns upon local driving, and reciprocally such tunable response upon far field driving can be viewed as a poor mans version of coherent control, where the phase and amplitude of driving serve to optimize hot spots or directionality\cite{Kao2010,Stockman2002,Ni2012}.  We envision that the localization and directionality could be further optimized in future experiments by using ultrafast fs waveguide excitation and pulse shaping strategies\cite{Aeschlimann2007}. As a possible application one can envision the use of this platform of waveguide-addressable spatially tunable hot spots for, for instance spatially cross correlated spectroscopies, such as fluorescence correlation spectroscopy\cite{Aouani2011}.\par

\subsection{In-coupling by a Yagi-Uda antenna}
\begin{figure}[!htb]
  \includegraphics[width=8cm]{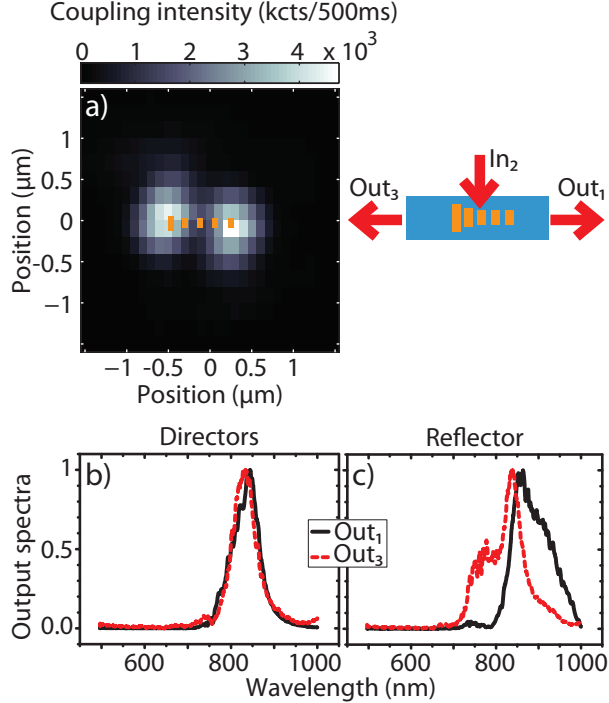}\\
  \caption{a) Confocal image of the waveguide obtained by scanning a focused spot over the Yagi-Uda antenna on top of a 1500~nm width waveguide and collecting spectra of light coupled into the waveguide at the waveguide end facet. The map is created by plotting the integrated count rate of the spectra taken at each position of the sample.  b) and c) Spectra acquired from the positions of maximum in-coupling located at the directors and reflector side of the antenna, when measured through channel~1 and channel~3, as depicted in the top right sketch of the experiment.
  These graphs show the different spectral behaviour of the different parts of the antenna, namely, when exciting the directors the same spectra emerge from both waveguide ends. In contrast when locally exciting the reflector and feed element the spectra coupled into both forward and backward guided mode are highly different. The observations show high-contrast unidirectional coupling into the waveguides that can be reversed by sweeping wavelength.}\label{YUmultielementAllPartB}
\end{figure}
As a final aspect of our experiment, we report on the in-coupling into the waveguide mode of a 1500~nm waveguide that can be achieved by raster scanning a focused spot over the antenna.  Our theory has shown that for in-coupling of plane waves a multiple of $\lambda_{waveguide}$ spacing is optimal so that a grating coupling effect aids in-coupling.  For diffraction limited illumination of just a few antenna elements, however, one might almost achieve a situation in which just one element of the antenna is excited. Instead of excitation by a molecule, excitation of just the plasmonic feed element by scattering can give rise to mimicking antenna directivity in a scattered signal, as realized by Kosako and Hoffman for antennas on an air-glass interface\cite{Kosako2010}. Therefore, we record scattering into the forward and backward waveguide direction as a function of where we illuminate the antenna with a tight focus. Surprisingly, as in scattering, two distinct zones of high in-coupling are found when we collect signal in both waveguide branches, and integrate over the full white light spectrum. Using spectrally resolved detection we assess wether these two zones of efficient in-coupling, one of which is at the feed element of the antenna, and the other of which is at the directors, are associated with the same or with different spectral features in the light that is coupled into both waveguide ends. In Fig.~\ref{YUmultielementAllPartB}b we plot the spectrum that is collected at both the forward, and the backward waveguide end when we excite the directors of the antenna. When the excitation spot is focused on the directors,  almost identical spectra emerge from both waveguide ends.  In stark contrast,  when the excitation spot is focused on the reflector side of the antenna, i.e., largely on the feed element, the spectra that emerge at the two waveguide ends are very different from each other.  At the end facet corresponding to the backward direction (reflector side of the antenna)  we obtain a spectrum that is significantly blue shifted from the spectrum retrieved at the the forward direction end facet.  The steep gradient in the spectra around 830~nm imply that it is possible to completely reverse the direction of the in-coupled beam that is launched into the waveguide with a very high contrast, simply by a small change in excitation wavelength.  This behavior is reciprocal to the strongly dependent receiver response of antenna arrays, that is expected to swap directionality as the excitation wavelength is swept through cutoff, phenomenon first reported for plasmonic antennas in experiments by de Waele et al.\cite{Waele2007}. To conclude we demonstrate that for localized excitations created by a focused spot at the feed element, Yagi-Uda antennas present directional coupling into the 1D waveguide. This in-coupling process is spectrally centered around $\sim830$~nm which is the same resonance frequency as for the resonant scattering of guided modes into free space.
When examining the in-coupling count rates in Fig.~\ref{YUmultielementAllPartB}a for the Yagi-Uda antenna we see that the in-coupling for the Yagi-Uda is around three times more efficient than for the single element antenna.
\subsection{Outlook}

To conclude, we have fabricated plasmonic antennas precisely aligned to dielectric waveguides, and quantified their properties for applications in waveguide-integrated plasmonics. As a first step, we have quantified how single plasmonic dipole antennas couple to waveguide modes, in particular quantifying how strongly, and into which directions, antennas outcouple waveguide modes. Conversely, we have shown that a single plasmon antenna can already couple up to 20\% of a diffraction limited input beam into the waveguide mode. Secondly, we have demonstrated how one can use the single dipole antenna as building block of a rational design strategy for multi-element antennas that derive functionality from a phased-array coherent response to driving by the waveguide mode. In particular, we have demonstrated that waveguide-coupled Yagi-Uda antennas provide a platform for waveguide addressable spatially tunable hot spots, that can for instance be used as programmable hot spots of pump light for spatially cross correlated spectroscopies. Conversely, the antennas can provide strong directionality, notably allowing to couple a local driving field efficiently and unidirectionally into the waveguide. While our experimental demonstration of this unidirectional coupling used an external laser scattered off the antenna, our calculations show that this functionality will directly extend to fluorophores, Thereby, waveguide-hybridized plasmon array antennas are a highly promising platform for many applications. For instance, one can efficiently collect all the fluorescence of single
fluorophores directly through a waveguide. The combination with the programmable hot spots of pump light that can be generated, makes this platform highly attractive for making optofluidic lab-on-chip devices that have entirely on chip integration of driving and readout for advanced fluorescence assays at single molecule levels. Also, we envision that hybrid systems of plasmon antennas and dielectric waveguides may outperform proposed plasmon quantum optics schemes\cite{Akimov2007,Falk2009}. While plasmonics offer very high light-matter interaction strength for coupling to localized emitters that act as qubits, the structures with the highest interaction strength are usually least suited as waveguides for transport, as losses are high. We propose that the combination of antennas and dielectric waveguides allows one to combine low loss transport as photons that are converted back and forth to plasmons only exactly where needed, i.e., at an antenna surrounding an emitter.

\begin{acknowledgement}
We thank L. Langguth, M. Frimmer, H. Schokker, A. Mohtashami, I. Sersic, G. Vollenbroek  and L. Huisman for experimental help and fruitful discussions.
This work is part of the research program of the ``Stichting voor Fundamenteel
Onderzoek der Materie (FOM)'', which is financially supported
by the ``Nederlandse Organisatie voor Wetenschappelijk
Onderzoek (NWO)''. This work is supported by NanoNextNL, a micro and nanotechnology consortium of the Government of the Netherlands and 130 partners.
\end{acknowledgement}

\section{Methods}

\subsection{Microscope setup}
In order to study antennas coupled to dielectric waveguides, we employ a setup that combines a fiber-coupled end-fire setup with a confocal microscope as seen in figures \ref{setupandsample}a~and~\ref{setupandsample}b. The setup can be used in two configurations, as further highlighted in the sketches presented on the left side of the figures. In the first configuration of the setup shown in  Fig.\ref{setupandsample}a light is coupled from  one end facet into the waveguide using a cleaved fiber (Nufern S630\_HP) that carries  excitation light from a Fianium supercontinuum light source (SC-450-PP, with the spectrum after the fiber ranging from 650 to 900 nm, max. power at 725~nm of 0.680~mW when measured through bandpass filter 700~nm FWHM 50~nm). Light is coupled into the waveguide and transmitted into a second fiber for spectral analysis on an Avantes peltier cooled Si CCD array spectrometer (AvaSpec-2048TEC-USB2-2).
To quantify the scattered light spectrally, spatially and in terms of wave vector content, a home built microscopy system is placed with its optical axis perpendicular to the sample substrate. We use a Olympus 100x, NA 0.95 M Plan IR objective to collect the scattered light, which is then directed through a tube lens to a CCD camera (The Imaging Source DMK21AU04) for imaging, or to a Thorlabs galvo scanner system. This galvo system scans the scattered light collected  from the sample plane over a 50~$\mu$m core multimode fiber which acts as a confocal pinhole (sample-to-fiber magnification 228 times
). This fiber brings out-of-plane scattered light onto a second channel of the same Avantes spectrometer. This confocal scanning configuration for out-of-plane scattering allows us to retrieve images of the sample as well as the spectral content of light scattered from different parts of the antenna. As further functionality, we can flip in a so-called Fourier or Bertrand lens that allows conoscopic imaging. In other words, when flipping in the  Fourier lens we retrieve the intensity distribution of scattered light over all wave vectors in the objective NA, essentially through imaging the back focal plane (BFP) of the imaging objective \cite{sersic2011, Aouani2011, LeThomas2007, Drezet2008,Randhawa:10,alaverdyan2009,Huang2008,Lieb:04,patra2005}. By using a pinhole system at a distance equal to the focal distance $f_{Fourier}$ from the Fourier lens we spatially filter the scattered light prior to wave vector imaging, so that we collect radiation patterns only from those  parts of the sample that we are interested in, namely the antennas. The Fourier image can again be collected panchromatically on the CCD, or through the galvo scanning mirrors by the fiber, which allows us to spectrally resolve the differential scattering cross section. To conclude, with this configuration of the setup we can study the effect of the antenna on the waveguide transmission (channel 3 in Fig.\ref{setupandsample}a) and scattering of the antenna into the air side (channel 2 in Fig.\ref{setupandsample}a). Given the thickness of the quartz substrate used for the samples ($\sim600$~$\mu m$), a home made solid immersion lens system (SIL) was required in order to also access scattering into the substrate side  (depicted as   channel 4 in Fig.\ref{setupandsample}a), as the Olympus objective lacks the required working distance. This SIL system  that employs a BK7 glass  hemisphere of diameter 2~mm, allowed us also to collect light that was scattered by the antennas into angles that exceed the total internal reflection angle of the substrate, however, only with spherical and chromatic imaging aberrations too large to allow diffraction limited and Fourier imaging.\par
The second configuration of the microscope in Fig.\ref{setupandsample}b is designed to study the converse interaction, i.e., rather than coupling in through the waveguide and collecting scattered light, we  study how light coming from free space (channel 2 in Fig.\ref{setupandsample}b) is coupled into the waveguide mode. This configuration is achieved by swapping the spectrometer-coupled detection fiber  that is placed after the galvo system with the combination of a pinhole system and free space collimated supercontinuum light from the Fianium source. The light scattered into the waveguide and detected through the fibers at the end facets is  sent to the spectrometer to quantify the  "forward" and "backward" waveguide incoupling spectra.  In absence of the  Bertrand lens, we  couple in light locally using real space focusing at the diffraction limit. When we flip the Bertrand lens in so that the incident beam is focused in the objective back aperture, we couple light in over a large area, yet at a well-defined incident angle that can be freely varied over the entire objective NA.

\subsection{Sample fabrication}
As we ultimately target visible light spectroscopy applications, we consider silicon nitride waveguides.  Fused silica wafers (n=1.45) of 100~mm diameter were covered with 100~nm thick Si$_{3}$N$_{4}$ using a LPCVD process.  This process [Lionix BV, The Netherlands] ensures low loss Si$_{3}$N$_{4}$ at manageable stress levels for postprocessing.  In order to define 1D waveguide ridges, we perform e-beam lithography using a Raith e-line machine. The waveguides together with positioning markers that are used at a later stage were defined in MaN2403 negative resist (200 nm thickness) with an electron beam lithographic step [dose 235~$\mu$C/cm$^2$, 35~nm spot size, current 0.14~nA and a fixed beam movable stage (FBMS) step size of 0.01~$\mu$m].  The pattern was then transferred into the Si$_{3}$N$_{4}$ by dry etching (Oxford Plasmalab, 50~sccm CHF$_{3}$ and 5~sccm O$_{2}$, 100~W forward RF power, 5~min etching time). In a second electron beam lithography step the antennas were defined on top of the waveguides, using alignment markers fabricated in the Si$_{3}$N$_{4}$ for precise positioning. In this step ZEP-520A positive resist (125~nm thick, exposed with a line dose of 200~pA s/cm, 29~nm spot size, current 0.03~nA)  was used to define a liftoff mask for thermal vapor deposition of gold. To mitigate the very poor adhesion of gold on Si$_{3}$N$_{4}$, in the evaporation step we first deposited a thin chromium adhesion layer of $\sim$3~nm, prior to the deposition of $\sim$30~nm of gold. A typical  final result is shown in Fig.\ref{setupandsample}d.\par
In this paper we discuss  two different types of antennas, namely single rod 100~nm long antennas and so called Yagi-Uda antennas as described in reference\cite{Curto2010}. One such Yagi-Uda antenna is shown in Fig.\ref{setupandsample}c~and~d. This antenna is composed of 5 rod shaped elements, 3 directors, 1 feed element and 1 reflector, with lengths of 115~nm 120~nm and 180~nm, respectively. The measured center-to-center distances between the elements are 170~nm between reflector and feed element, and  183~nm between all other neighboring plasmonic rods. The width of the elements is 65~nm and the total length of the antenna is $\sim$790~nm. While we have studied antennas on various waveguide widths, all the data presented here are for waveguide widths of 1000~nm and 1500~nm. The strip height is 100~nm. We estimate the electron beam alignment accuracy of antennas to waveguides to be $\sim$40~nm, i.e., far below any typical feature of the waveguide mode structure.

\subsection{Spectra normalization}
A particularly difficult problem is how to quantitatively normalize  the spectrum of light scattered by the antennas to the spectrum that is offered through the waveguide to the antenna. The only possible references we have access to are the spectra measured in transmission through nominally identical blank waveguides (i.e., without antennas) as a measure for the incident spectrum, and  spectra obtained from scattering centers that appear comparatively close to the antenna, due to  roughness of the waveguide.  In the first case, artifacts may occur due to the fact that the spectrum may vary between alignments and between waveguides, due to chromatic effects in coupling to the waveguide, and for the  1~cm long waveguides due to the integrated effect of  unexpected small defects and impurities that change the spectrum along the length of the waveguide.  In the second approach, the advantage is  that spectra are taken from a region very close to the structure. However,  one here relies on the assumption that the scattering centers have no strong frequency dependence, and one does not obtain a quantitative signal strength comparison,  as opposed to when using waveguide transmission. In practice no large difference between the two approaches is found when spectrally locating the resonance. Here we present data using the second method (normalization to nearby scattering centers), preferring spectral fidelity over an absolute scale.

\bibliography{WaveguideYagiUda}

\end{document}